\documentstyle[prl,twocolumn,aps]{revtex}
\input{epsf}

\def\beq{\begin{eqnarray}}
\def\eed{\end{eqnarray}}
\begin{document}
\tightenlines
\draft

\title{The effects of Zn Impurity on the Properties of Doped Cuprates in the Normal State}
\author{Yun Song}
\address{Department of Physics, Beijing Normal University, Beijing
100875, China}

\date{\today }

\maketitle

\begin{abstract}
We study the interplay of quantum impurity, and collective spinon
and holon dynamics in Zn doped high-T$_c$ cuprates in the normal
state. The two-dimensional t-t$^{\prime}$-J models with one and a
small amount of Zn impurity are investigated within a numerical
method based on the double-time Green function theory. We study
the inhomogeneities of holon density and antiferromagnetic
correlation background in cases with different Zn concentrations,
and obtain that doped holes tend to assemble around the Zn
impurity with their mobility being reduced. Therefore a bound
state of holon is formed around the nonmagnetic Zn impurity with
the effect helping Zn to introduce local antiferromagnetism around
itself. The incommensurate peaks we obtained in the spin structure
factor indicate that Zn impurities have effects on mixing the
q=($\pi$, $\pi$) and q=0 components in spin excitations.
\end{abstract}
\vskip1pc

\narrowtext

The effects of divalent transition metal Zn substitute for Cu in
CuO$_{2}$ plane present much valuable information in understanding
the mechanism of high temperature superconductors. Zn$^{2+}$ has a
closed $d$ shell with spin $s=0$, and acts as a very strong
scattering center. As a result, the spin configurations and the
electronic structures around the nonmagnetic impurity Zn are
strongly disrupted in both the normal state and superconducting
state. Below $T_{c}$, there have been many
experimental\cite{Pan,Sidis,Hudson,Yazdani,Mahajan} and
theoretical\cite{Polkovnikov,Balatsky,Jian,Tsuchiura,Xiang}
investigations to discover how the $d$-wave superconductivity is
destroyed and what is the microscopic mechanism behind. As the
normal state properties are more fundamental, some
experimental\cite{Alloul,Bobroff,Julien,MacFarlane,Yamagata} and
theoretical\cite{Nagaosa,Bulut,Kruis} works have been done to
study the effects of nonmagnetic impurity Zn on the properties of
normal state.  It is believed that these studies are of great help
for understanding the peculiar behaviors of normal state.
Moreover, they shed light on the recent striking issue of the
normal state pseudogap \cite{Tallon,Chattopadhyay}. So far, there
has been no clear picture of how the impurity interplays with the
strong correlation background. It still remains unanswered why the
Zn impurity produced very strong scatter and how the impurity band
forms in real space.

In this paper, we perform numerical calculation to study the
effects of Zn impurity on its surrounded Cu ions in the normal
state. We want to find out how the Zn impurity influences the hole
distribution and AF correlation background, which may bring about
a better understanding of the fundamental relation between spin
and hole. We start from the two-dimensional (2D) t-t$^{\prime}$-J
model and use fermion-spin theory \cite{Feng}. Fermion-spin theory
is based on the charge-spin separation, in which the single
occupied constrain of t-t$^{\prime}$-J model could be treated
properly even in the mean field approximation. Within an improved
Green function theory \cite{Song}, we perform numerical
calculation for cases with only one Zn impurity and a small amount
of Zn impurity, and the effect of Zn concentration on some
properties of the normal state are discussed.

The essential physics of high-$T_{c}$ cuprates is well described
by the t-J model on a square lattice. In the condition that Zn
substitutes Cu in CuO$_{2}$ plane, we can model Zn impurity as
vacant site, which has no coupling with surrounded Cu sites. We
add the next-nearest-neighbor hopping term in our model to
reproduce the realistic band structure, and start our study from
the following Hamiltonian
\begin{eqnarray}
H&=&-t\sum_{\langle i,j\rangle \neq l,\sigma }(C_{i\sigma
}^{\dagger }C_{j\sigma }+h.c.)-t^{\prime}\sum_{\langle
i,i^{\prime}\rangle \neq l,\sigma }(C_{i\sigma }^{\dagger
}C_{i^{\prime}\sigma
}+h.c.)\nonumber\\
&&-\mu \sum_{i,\sigma }C_{i\sigma }^{\dagger }C_{i\sigma }
+J\sum_{\langle i,j\rangle \neq l}\mathbf{S}_{i}\cdot \mathbf{S}
_{j},
\end{eqnarray}
where $\langle i,j\rangle$ and $\langle i,i^{\prime}\rangle$ mean
the summations over nearest-neighbor (NN) and
next-nearest-neighbor (NNN) pairs respectively, and $l$ represents
the site occupied by Zn impurity.  In our model the direct hopping
among a Zn impurity and its surrounded Cu sites is forbidden. In
addition, to eliminate the doubly occupied sites of Cu ion, we
introduce constraint
$\sum_{\sigma}C_{i\sigma}^{\dagger}C_{i\sigma}\leq 1$ for each Cu
site. Also we introduce
$\sum_{\sigma}C_{l\sigma}^{\dagger}C_{l\sigma}=2$ for Zn$^{2+}$
ion since it has a closed $d$ shell. Therefore, the total number
of electrons satisfies
$\sum_{i\sigma}C_{i\sigma}^{\dagger}C_{i\sigma}=N-N_{h}+N_{Zn}$
with $N_{h}$ and $N_{Zn}$ representing the number of hole and Zn
impurity, respectively. As the strong electron correlation
manifests itself by the local constraint, the key issue is how to
treat the constraint properly.

Here we study the t-t$^{\prime}$-J model within the fermion-spin
theory\cite{Feng} based on the charge-spin separation. We
introduce $C_{i\uparrow}= h_{i}^{\dagger}S_{i}^{-}$ and
$C_{i\downarrow}=h_{i}^{\dagger} S_{i}^{+}$, where the spinless
fermion operator $h_{i}$ describes the charge (holon) degrees of
freedom, while the pseudospin operator $S_{i}$ describes the spin
(spinon) degrees of freedom. Thus the low energy behavior of the
t-t$^{\prime}$-J model (1) can be written as
\begin{eqnarray}
H &=&-t\sum_{\langle i,j\rangle \neq l
}(h_{i}h_{j}^{\dagger}+h_{j}h_{i}^{\dagger})
(S_{i}^{+}S_{j}^{-}+S_{i}^{-}S_{j}^{+})\nonumber\\
&&-t^{\prime}\sum_{\langle i,i^{\prime}\rangle \neq l
}(h_{i}h_{i^{\prime}}^{\dagger}+h_{i^{\prime}}h_{i}^{\dagger})
(S_{i}^{+}S_{i^{\prime}}^{-}+S_{i}^{-}S_{i^{\prime}}^{+})
\nonumber\\
&&+\mu\sum_{i\neq l}h_{i}^{\dagger}h_{i}+ \sum_{\langle i,j\rangle
\neq l}J^{{\rm eff}}_{i,j}{\bf S}_{i}\cdot {\bf S}_{j},
\end{eqnarray}
where $J^{{\rm
eff}}_{i,j}=[(1-n^{h}_{i})(1-n^{h}_{j})-\phi^{2}_{i,j}]J$ with
$n^{h}_{i}$ representing the hole concentration at site $i$ and
$\phi _{i,j} =\langle h_{i}^{\dagger} h_{j}\rangle$ being the
order parameter of holon. Here we also introduce spinon
correlation functions $\chi_{ij}=\langle
S_{i}^{-}S_{j}^{+}\rangle$ and $\chi_{ij}^{z}=\langle
S_{i}^{z}S_{j}^{z}\rangle$.

We introduce three double-time Green functions
\begin{eqnarray}
G_{h}(i-j,~\tau-\tau^{\prime })&=&-i\theta (\tau-\tau^{\prime
})\langle [h_i(\tau);~h_{j}^{+}(\tau ^{\prime })]\rangle
\nonumber\\
&\equiv &\langle \langle h_{i}(\tau );
h_{j}^{+}(\tau ^{\prime })\rangle \rangle \nonumber\\
D_{s}(i-j,~\tau-\tau^{\prime })&=&-i\theta (\tau-\tau^{\prime
})\langle [S_{i}^{+};~S_{j}^{-}(\tau ^{\prime })]\rangle
\nonumber\\
&\equiv &\langle \langle S_{i}^{+}(\tau );S_{j}^{-}(\tau ^{\prime
})\rangle \rangle  \nonumber \\
D_{s}^{z}(i-j,~\tau-\tau^{\prime
})&=&-i\theta (\tau-\tau^{\prime })\langle
[S_{i}^{z};~S_{j}^{z}(\tau ^{\prime })]\rangle
\nonumber\\
&\equiv &\langle \langle S_{i}^{z}(\tau ); S_{j}^{z}(\tau ^{\prime
})\rangle \rangle ,
\end{eqnarray}
where $G_{h}$ describes the behaviors of holon, and $D_{s}$ and
$D_{s}^{z}$ describe the behaviors of spinon. Since the lattice
translational invariance is not presented in cases with Zn
impurities, we evaluate the equations of motion of the above
Green's functions in real space. The double-time Green function
$\langle \langle A;~B \rangle \rangle$ satisfies
\begin{eqnarray}
\omega \langle \langle A;~B\rangle \rangle _\omega =\langle
[A,~B]_{\mp}\rangle _\omega +\langle \langle [A,~H];~B\rangle
\rangle _\omega ,
\end{eqnarray}
thus we could obtain the equations of motion of $G_{h}$
\begin{eqnarray}
(\omega-\mu)G_{h}(i-j)_{\omega }-2t\sum_\eta \chi _{i,i+\eta
}G_{h}(i+\eta-j)_{\omega }\nonumber\\
-2t^{\prime } \sum_\tau \chi
_{i,i+\tau }G_{h}(i+\tau-j)_{\omega }=\delta (i-j).
\end{eqnarray}
We introduce $\widetilde{G}_{h}$, a $N^2\times N^2$ elements
matrix, to express the holon Green functions for a square lattice
with $N\times N$ sites. And we could rewrite Eq. (5) as
\begin{eqnarray}
(\omega -\mu)\widetilde{G_{h}}-\widetilde{h}\
\widetilde{G_{h}}=\widetilde{I},
\end{eqnarray}
where matrix $\widetilde{h} $ is decided by the NN and NNN spinon
correlation functions, and $\widetilde{I} $ is an identity matrix.

Based on Eq. (3) and (4), we also obtain the equations of motion
of spinon Green functions $D_s$ and $D_{s}^{z}$
\begin{eqnarray}
\omega D_{s}(i-j)_{\omega }&=&2\sum_\eta J^{eff}_{i,i+\eta }
\{\epsilon
_{i,i+\eta}F_{1}(i,i+\eta;~j)_{\omega}\nonumber\\
&&-F_{1}(i+\eta ,i ;~j)_{\omega}\}
\nonumber\\
&&+8t^{\prime} \sum_\tau
\phi_{i,i+\tau}F_{1}(i,i+\tau;~j)_{\omega}
\nonumber\\
\omega D^{z}_{s}(i-j)_{\omega }&=&\sum_\eta J^{eff}_{i,i+\eta }
\epsilon_{i,i+\eta}
\{F_{2}(i,i+\eta;~j)_{\omega}\nonumber\\
&&-F_{2}(i+\eta,i;~j)_{\omega}\}
\nonumber\\
&&+4t^{\prime} \sum_\tau \phi_{i,i+\tau}
\{F_{2}(i,i+\tau;~j)_{\omega}\nonumber\\
&&-F_{2}(i+\tau ,i;~j)_{\omega}\},
\end{eqnarray}
where $\epsilon
_{i,i+\eta}=1+\frac{4t\phi_{i,i+\eta}}{J_{i,i+\eta}^{eff}}$.
$F_{1}$ and $F_{2}$ are the second-order spinon Green functions
which are defined as
\begin{eqnarray}
F_{1}(i,l;~j)_{\omega}&=&\langle \langle S_{i}^{z}
S_{l}^{+};~S_j^{-}\rangle \rangle _\omega\nonumber\\
F_{2}(i,l;~j)_{\omega}&=&\langle \langle S_i^{+}
S_{l}^{-};~S_j^{z}\rangle \rangle _\omega .
\end{eqnarray}
Going a step further, we establish the equations of motion of the
second-order spinon Green functions
\begin{eqnarray}
\omega F_{1}(i,l;~j)_{\omega}&=&
2\chi_{i,l}^{z}\delta(l-j)-\chi_{i,l}\delta(i-j)\nonumber\\
&+&\langle \langle \{ \sum_{\eta} [ 2J_{l,l+\eta}^{eff} (\epsilon
_{l,l+\eta} S_{i}^{z}S_{l}^{z}S_{l+\eta}^{+}
-S_{i}^{z}S_{l+\eta}^{z}S_{l}^{+})\nonumber\\
&+&J_{i,i+\eta}^{eff}\epsilon_{i,i+\eta}
(S_{i}^{+}S_{i+\eta}^{-}S_{l}^{+}-S_{i+\eta}^{+}S_{i}^{-}S_{l}^{+})]
\nonumber\\
&+&4t^{\prime} \sum_{\tau} [ \phi_{i,i+\tau}
(S_{i}^{+}S_{i+\tau}^{-}S_{l}^{+}-
S_{i+\tau}^{+}S_{i}^{-}S_{l}^{+})\nonumber\\
&+&2\phi_{l,l+\tau}S_{i}^{z}S_{l}^{z}S_{l+\tau}^{+} ] \}
;S_{j}^{-} \rangle \rangle _{\omega}
\nonumber\\
\omega F_{2}(i,l;~j)_{\omega}&=&
\chi_{i,l}\delta(l-j)-\chi_{i,l}\delta(i-j) \nonumber\\
&+&\langle \langle \{ \sum_{\eta} [2J_{l,l+\eta}^{eff}
(S_{i}^{+}S_{l}^{-}S_{l+\eta}^{z}
-\epsilon_{l,l+\eta}S_{i}^{+}S_{l+\eta}^{-}S_{l}^{z}) \nonumber\\
&+&2J_{i,i+\eta}^{eff} (\epsilon_{i,i+\eta}
S_{i+\eta}^{+}S_{l}^{-}S_{i}^{z}-S_{i}^{+}S_{l}^{-}S_{i+\eta}^{z})]
\nonumber\\
&+&8t^{\prime} \sum_{\tau} [\phi_{i,i+\tau}
S_{i+\tau}^{+}S_{l}^{-}S_{i}^{z}\nonumber\\
&&-\phi_{l,l+\tau}
S_{i}^{+}S_{l+\tau}^{-}S_{l}^{z}] \};~S_{j}^{z} \rangle \rangle
_{\omega}.
\end{eqnarray}
To the third-order spinon Green functions in the right hand side
of Eq. (9), we perform the improved decoupling scheme as described
in Ref. 22, for example
\begin{equation}
\langle \langle S_{i}^{z}S_{l}^{z}S_{l+\eta}^{+};~S_{j}^{-}\rangle
\rangle \rightarrow \alpha_{i} \langle S_{i}^{z}S_{l}^{z}\rangle
\alpha_{l} \langle \langle S_{l+\eta}^{+};~S_{j}^{-}\rangle
\rangle .
\end{equation}
Therefore, the second-order spinon Green functions $F_{1}$ and
$F_{2}$ can be expressed by the Green functions $D_{s}$ and
$D_{s}^{z}$ as
\begin{eqnarray}
\omega F_{1}(i,l;~j)_{\omega}&=&\Gamma_{1}^{0}+
\Gamma_{1}^{1}D(i-j)+\Gamma_{1}^{4}D(l-j) \nonumber\\
&+&\sum_{\eta} \{ \Gamma_{1}^{2}D(i+\eta-j)
+\Gamma_{1}^{3}D(l+\eta-j) \} \nonumber \\
&+&\sum_{\tau} \{ \Gamma_{1}^{5}D(i+\tau-j)
+\Gamma_{1}^{6}D(l+\tau-j) \} \nonumber\\
\omega F_{2}(i,l;~j)_{\omega}&=&\Gamma_{2}^{0}+
\Gamma_{2}^{1}D^{z}(i-j)
+\Gamma_{2}^{4}D^{z}(l-j)\nonumber\\
&+&\sum_{\eta} \{ \Gamma_{2}^{2}D^{z}(i+\eta-j)
+\Gamma_{2}^{3}D^{z}(l+\eta-j) \}, \nonumber \\
\end{eqnarray}
where
\begin{eqnarray}
\Gamma_{1}^{0} &=&2\chi_{i,l}^{z} \delta (l-j) -\chi_{i,l} \delta
(i-j)\nonumber\\
\Gamma_{1}^{1} &=&\sum_{\eta}J_{i,i+\eta}^{eff}
\epsilon_{i,i+\eta} \alpha_{i+\eta}\chi_{i+\eta,l}\alpha_{l}
\nonumber \\
&&+4t^{\prime} \sum_{\tau} \phi_{i,i+\tau}
\alpha_{i+\tau}
\chi_{i+\tau,l} \alpha_{l}\nonumber\\
\Gamma_{1}^{2} &=&-J_{i,i+\eta}^{eff} \epsilon_{i, i+\eta}
\alpha_{i} \chi_{i,l} \alpha_{l}\nonumber\\
\Gamma_{1}^{3} &=&2J_{l,l+\eta}^{eff} \epsilon_{l,l+\eta}
\alpha_{i} \chi_{i,l}^{z} \alpha_{l}\nonumber\\
\Gamma_{1}^{4} &=& -2\sum_{\eta} J_{l,l+\eta}^{eff} \alpha_{i}
\chi_{i,l+\eta}^{z} \alpha_{l+\eta}\nonumber\\
\Gamma_{1}^{5} &=&-4t^{\prime} \phi_{i,i+\tau} \alpha_{i}
\chi_{i,l} \alpha_{l}\nonumber\\
\Gamma_{1}^{6} &=&8t^{\prime} \phi_{l,l+\tau} \alpha_{i}
\chi_{i,l}^{z} \alpha_{l}\nonumber\\
\Gamma_{2}^{0} &=&\chi_{i,l} \delta (l-j) - \chi_{i,l} \delta
(i-j)\nonumber\\
\Gamma_{2}^{1} &=&2\sum_{\eta} J_{i,i+\eta}^{eff}
\epsilon_{i,i+\eta} \beta_{i+\eta} \chi_{i+\eta,l} \beta_{l}
\nonumber \\
&&+8t^{\prime} \sum_{\tau} \phi_{i,i+\tau} \beta_{i+\tau}
\chi_{i+\tau,l} \beta_{l}\nonumber\\
\Gamma_{2}^{2} &=&-2J_{i,i+\eta}^{eff} \beta_{i} \chi_{i,l}
\beta_{l}\nonumber\\
\Gamma_{2}^{3} &=&2J_{l,l+\eta}^{eff} \beta_{i} \chi_{i,l}
\beta_{l}\nonumber\\
\end{eqnarray}
and \begin{eqnarray} \Gamma_{2}^{4} &=&-2\sum_{\eta}
J_{l,l+\eta}^{eff} \epsilon_{l,l+\eta} \beta_{i} \chi_{i,l+\eta}
\beta_{l+\eta}\nonumber \\
&&-8t^{\prime} \sum_{\tau} \phi_{l,l+\tau} \beta_{i}
\chi_{i,l+\tau} \beta_{l+\tau}.\nonumber
\end{eqnarray}
 We also
introduce two $N^2\times N^2$ elements matrices
$\widetilde{D}_{s}$ and $\widetilde{D}_{s}^{z}$ to express the
spion Green functions. Based on Eq. (7) and (11), we could obtain
that
\begin{eqnarray}
\omega ^2\widetilde{D_s}-\widetilde{S}\
\widetilde{D_s}&=&\widetilde{C_s}\nonumber\\
\omega ^2\widetilde{D^{z}_s}-\widetilde{S^{z}}\
\widetilde{D^{z}_s}&=&\widetilde{C^{z}_s},
\end{eqnarray}
where matrices $\widetilde{S} $, $\widetilde{S^{z}} $,
$\widetilde{C_{s}}$ and $\widetilde{C_{s}^{z}}$ are decided by the
density of holon $n_i$, and the spin correlation functions.

We establish the self-consistent equations based on Eq. (6) and
(12) to determine the correlation functions of holon and spinon,
and also the vertex correction parameters. Under the periodic
boundary conditions, we have performed numerical calculations for
$16 \times 16$ and $20 \times 20$ lattices with different Zn
concentrations respectively. Based on the experimental results of
BSCCO near optimal doping \cite{Pan,Tsuchiura}, the parameters of
t-t$^{\prime}$-J model are taken as $t/J=2.5$ and
$t^{\prime}/t=-0.4$ in our calculations.

Firstly, we study the $20\times 20$ lattice with only one Zn
impurity in the optimally doped region ($\delta_{h}=0.15$). The
spacial distribution of holon density is calculated and our
numerical results are shown in figure 1. We find that the holon
density closet to the Zn impurity oscillates strongly, and the
fluctuation diminishes rapidly away from the Zn impurity. The
maximum density is obtained at sites two lattice distances away
from the Zn impurity, and is about $15\%$ higher than the minimum
density. Our numerical results suggest that doped holes form a
local region around the Zn impurity, whose size is about eight
lattice cells as shown in figure 1. We also obtain that the
magnetic modification introduced by Zn is mainly in the vicinity
of Zn impurity. The NN correlation functions near the isolated
nonmagnetic impurity in the undoped case have been discussed
carefully in Ref. 22. We have found that Zn impurity strongly
modifies the spin excitations, especially the magnetic properties
of the neighbor Cu. As a result, the AF correlation functions at
the bonds closed to the impurities are enhanced. In the optimally
doped region, we also obtain that Zn impurity enhances the NN AF
correlation functions of spinons close to it.  Moreover, The doped
holes have the effect to strengthen the AF correlations near the
Zn impurity. The $^{63}$Cu NMR study of
YBa$_2$(Cu$_{0.99}$Zn$_{0.01}$)$_3$O$_{6.7}$ also find that the AF
correlations are enhanced, not destroyed, around Zn impurities
\cite{Julien}. Since the quantum fluctuation of spinons close to
the nonmagnetic impurity is reduced obviously, we can divided the
system into strong AF correlation region and weak AF correlation
region. The tendency of doped holes to assemble around the Zn
impurity could rationalize anomalous charge localization effect,
and the mobility of those holons closed to the Zn impurity could
also be reduced. Therefore, a bound state of holon
\cite{Poilblanc} is formed around the Zn impurity. Bound state of
impurity in normal state has also been predicted by the
self-consistent $T$ matrix approach\cite{Kruis}.

We also study the cases with several Zn impurities in the
optimally doped regime, and find that the holes play different
roles in the strong AF correlation region and the weak AF
correlation region. Our numerical results show that the AF
correlations of bonds far from the Zn impurity reduces remarkably
as hole is added into the systems. On the contrary, the bound
state of holon has the effect to enhance the AF correlations
around Zn impurity, and helps Zn to introduce local
antiferromagnetism around itself. As Zn concentration increases,
the CuO$_{2}$ plane becomes a inhomogeneous mixture of strong AF
correlation regions and weak AF correlation regions. And the doped
holes tend to assemble at the strong AF correlation regions.

We show the density of state of holon in cases with several Zn
impurities in figure 2. Jung $et$ $al$\cite {Jung} have examined
some samples and prove that Zn is uniformly distributed. In some
configurations when Zn is uniformly distributed, our calculation
show that due to the strong coupling between impurity and the
conduction band, the width of the holon band decreases as Zn
concentration increases.   We find that the inhomogeneity of the
spinon background increases the density of state of holon near the
fermi surface, and a resonance peak of impurity states is found to
get broader and stronger as Zn concentration increases. Inelastic
neutron scattering study for the optimal doped
La$_{1.85}$Sr$_{0.15}$Cu$_{1-y}$Zn$_{y}$O$_{4}$ indicates that a
new in-gap Zn impuriy state is introduced at low
temperature\cite{Kimura}. Nonmagnetic defect structures at the
surface has also been found to create localized low-energy
excitations in their immediate vicinity in
Bi$_2$Sr$_2$CaCu$_2$O$_8$ by performing low-temperature tunneling
spectroscopy measurements with a scanning tunneling
microscope\cite{Yazdani}. Our calculations show that the impurity
state can survive above T$_c$, which is in agreement with the
theoretical prediction \cite{Kruis} as well.

To study the effect of the Zn impurity on the spin background
around it, we introduce the spin structure factor
\begin{equation}
{\bf S}_{i}({\bf k})=\sum_{j}S_{i}^{z}S_{j}^{z}e^{{\bf
k}\cdot({\bf i}-{\bf j})}.
\end{equation}
Here $i$ represents Cu site around Zn impurity. Zn impurity is a
scatter which has a strong effect on the AF correlation
background\cite{Vajk}.  As Zn concentration increases, the
$CuO_{2}$ plane becomes a inhomogeneous mixture of strong AF
correlation regions and weak AF correlation regions. In figure 3,
the numerical results of the NN Cu sites around Zn impurity are
shown for $16\times 16$ lattices with different Zn concentrations.
We obtain that, in the pure case and Zn lightly doped case
($\delta_{Zn}\leq 0.01$), the spin excitations are dominated by a
magnetic resonance peak located at $Q_{AF}=(\pi, \pi)$. As the Zn
concentration increases, this peak decreases and there appear two
second-high incommensurate peaks as shown in figure 3, which
results from the mixing of $q=(\pi, \pi)$, $q=(\pi, 0)$ and $q=0$
components in spin excitations introduced by the strong impurity
scattering. In Bulut's study of susceptibility of Zn doped
high-$T_{c}$ superconductors, the similar behaviors are also
obtained \cite{Bulut}. In addition, we find that the distances of
the incommensurate peaks from $q=(\pi, \pi)$ increase with doping,
and these peaks become broad and weak in amplitude with the
increasing of Zn concentration. Meanwhile, as a result of the
increasing of the disorder introduced by the Zn substitution on
the Cu sites, the peak at $q=(\pi, \pi)$ decreases gradually as
the Zn concentration increases, and disappears when
$\delta_{Zn}\geq 0.1$.  Thus the result is consistent with
experimental results of Zn-doped high-$T_{c}$ cuprates
\cite{Alloul}.

In summary, we have studied the interplay between quantum
impurities, and collective spinon and holon dynamics in Zn-doped
cuprate in the normal state. Within a numerical method based on
the Green function theory, the inhomogeneities of holon density
distribution and antiferromagnetic correlation background in
two-dimensional t-t$^{\prime}$-J model with Zn impurities are
investigated. We obtain the real space shape of bound state of
holon surrounding the nonmagnetic Zn impurity. We also find that
the doped holes help Zn to introduce local antiferromagnetism
around itself. In the cases with a small amount of Zn impurities,
the influence of Zn impurity on the antiferromagnetic correlation
background is studied. The appearance of incommensurate peaks in
spin structure factor indicates that Zn impurity is a strong
scatter center, which has an effect on mixing the $q=(\pi, \pi)$,
$q=(\pi, 0)$ and $q=0$ components in spin excitations.

\acknowledgments The authors would like to thank Prof. Feng for
helpful discussions. This work was supported by the Grant from
Beijing Normal University.

\vglue 1.0cm
\begin{figure}
\epsfxsize=6cm \epsfysize=4cm \centerline{\epsffile{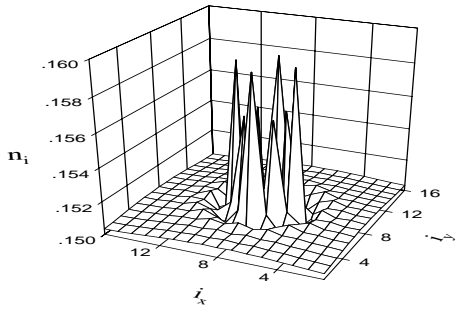}}
\vglue 0.5cm \caption{ The distribution of holon density in a $20
\times 20$ lattice with only one Zn impurity. }
\end{figure}

\vglue 1.0cm
\begin{figure}
\epsfxsize=6cm \epsfysize=4cm \centerline{\epsffile{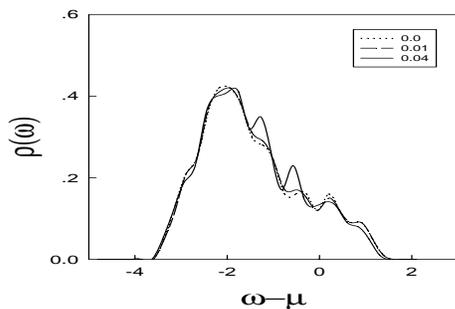}}
\caption{ The density of state of holon at $\delta_{Zn}$=0.0
(dotted line), 0.01(dashed line), and 0.04 (solid line). }
\end{figure}

\vglue 1.0cm
\begin{figure}
\epsfxsize=6cm \epsfysize=4cm \centerline{\epsffile{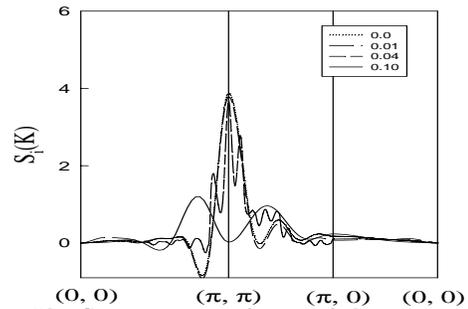}}
\caption{ The Spin structure factor of the nearest neighbor Cu
sites around Zn impurity for $16\times 16$ lattices with different
Zn concentrations. }
\end{figure}



\end{document}